# Are the Perseus-Pisces chain and the Pavo-Indus wall connected?

H. Di Nella[1,2], W.J. Couch[2], G. Paturel[1], Q.A. Parker[3]
[1] *Observatoire de Lyon, F69230 Saint-Genis Laval, FRANCE,*
[2] *School of Physics, University of New South Wales, Sydney, NSW 2052, AUSTRALIA*
[3] *Anglo-Australian Observatory, Coonabarabran, NSW 2357, AUSTRALIA*

**ABSTRACT**

In this paper we present a new redshift survey undertaken to address the rather interesting question that arises when all the currently available redshift information from the LEDA database is plotted in 'hypergalactic' coordinates. A significant empty region was found between the southern Pavo- Indus (PI) wall and the northern Perseus-Pisces (PP) chain. This survey tests the reality of this void which may simply reflect previous poor sampling of the galaxies in this region. Redshifts for a magnitude selected sample of 379 galaxies were obtained covering the four UKST/SERC survey fields: #537, #470, #346, #290 with $15.5 \leq B_T \leq 17.0$. All redshifts were obtained with the FLAIR multi-object spectroscopy system on the 1.2 m U.K. Schmidt Telescope at Siding Spring, Australia. Two highly significant density enhancements were found in the galaxy distribution at $133 Mpc$ and $200 Mpc$ ($H_0$=75 km.s$^{-1}$.Mpc$^{-1}$). We claim that no connexion exists between PP and PI. However, a southern extension of PP was detected and makes the total length of this chain of more than 150 Mpc.

## 1 INTRODUCTION

Over recent years, an increasingly detailed picture of the large- scale galaxy distribution of the local (d<200 Mpc) universe has emerged with the advent of large systematic redshift surveys (e.g. de Lapparent et al. 1986, Geller and Huchra 1989, Da Costa et al. 1988, Shanks et al. 1994).

Before the compilation of these large redshift surveys, astronomers mapped large-scale structures from the projected 2-D galaxy distribution based on either subjective visual studies (Lick catalogue, Shane & Wirtanen 1967) or more recently from objective machine-based scanning of the available Schmidt sky survey plates (Collins et al. 1992). However, rapid progress with redshift surveys led to more complex 3-D structures beeing discovered starting with the first discovery of a cell-like structure reported by Joeveer et al. (1978). Later Kirshner et al. (1981) found a very large empty region of galaxies: the so called Bootes void.

Anticipating the growth industry of redshift surveys and their use as a tool to describe the 3-D galaxy distribution, a collaboration between the Observatories of Lyon and Meudon was formed in 1983 to establish the first computer database to record all published information on galaxies: LEDA⋆. Around this time, the existence of superclusters, voids, filaments and sheets was being confirmed with typical feature scales of ∼ 50Mpc. In 1988, a Flamsteed's equal area projection of 58,000 galaxies, then available in the LEDA database, suggested the existence of a structure even larger than any previously seen (Bottinelli et al., 1986; Paturel et al., 1988). The structure seemed to connect several superclusters (Perseus- Pisces, Pavo-Indus, Centaurus and the Local Supercluster). The LEDA team calculated the pole of this flattened 'hypergalactic structure' to be about $l = 57$ deg; $b = 22$ deg in galactic longitude and latitude respectively. This most populated plane (MPP), also claimed by Tully in 1986 and 1987, is tilted by about ten degrees to the plane of the Local Supercluster (de Vaucouleurs 1953, 1956) $l = 47$ deg; $b = 6$ deg. Lynden-Bell (1991) also noticed this alignement of superclusters on the supergalactic plane via an equal area projection of galaxy positions.

Using the CfA1 redshift survey, De Lapparent, Geller and Huchra (1986) graphically confirmed the existence of voids and discovered a 'filament' in the Coma cluster region. Geller and Huchra (1989) showed that this filament was actually a sheet-like structure, that has become known as *the Great Wall*, similar to structures predicted in models by Zeldovich (1970). In the southern hemisphere, complementary redshift surveys were undertaken by da Costa et al. (1988) and Fairall et al., (1990). A compilation of available published Southern redshifts by Fairall and Jones, (1991) revealed another wall-like structure in the Pavo- Indus (PI) region. The Perseus-Pisces (PP) region was described either as a chain (Joever et al., 1978; Gregory, Thompson and Tifft, 1981) or a sheet-like structure (Haynes and Giovanelli 1986).

⋆ Lyon-Meudon Extragalactic Database. **telnet:** lmc.univ-lyon1.fr **login:** leda

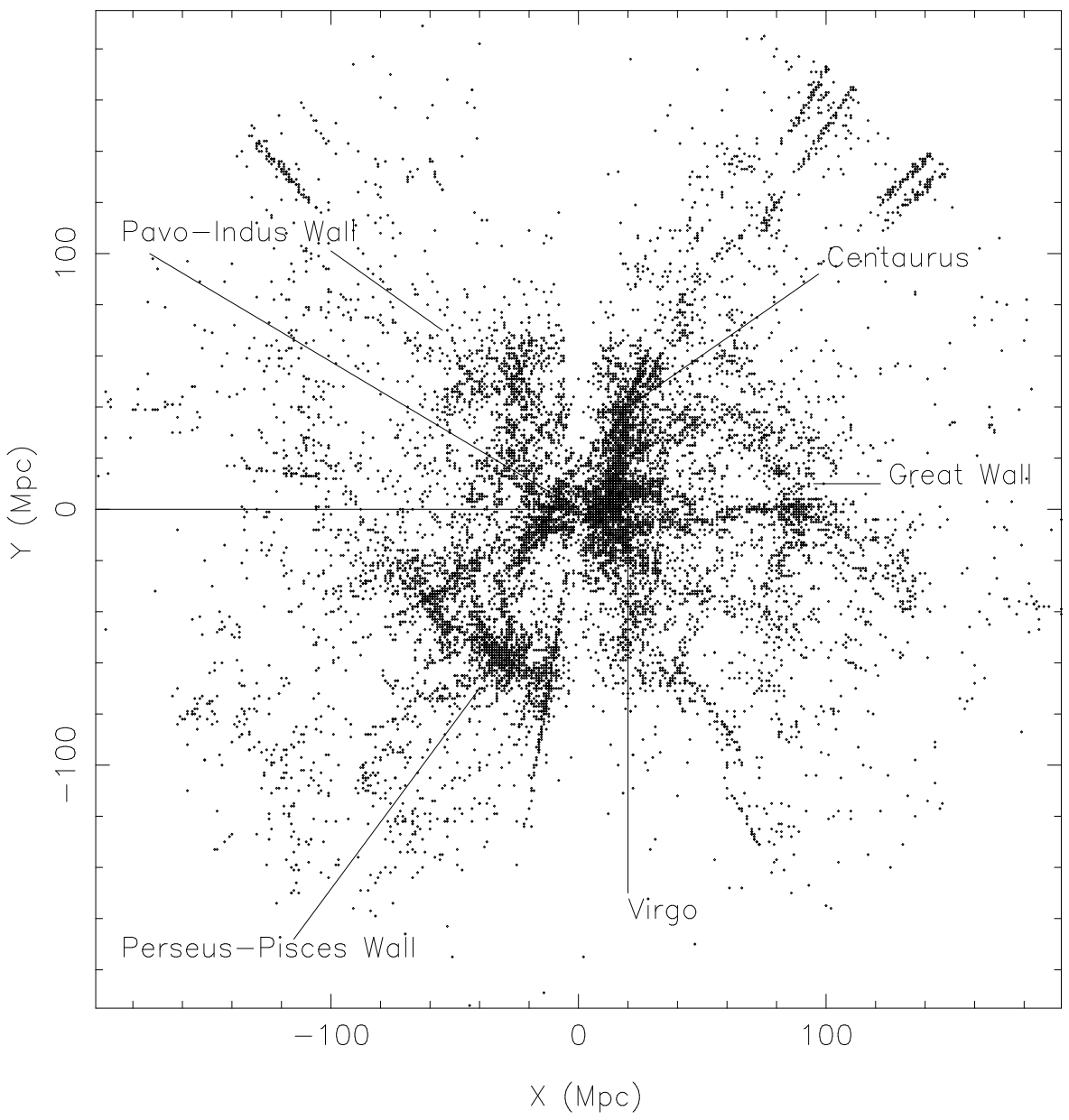

**Figure 1.** Distribution of galaxies from the LEDA sample in a face-on view of the so-called 'hypergalactic' plane (see text). The newly observed region is defined by the two continuous lines.

In the 12 years since the inception of LEDA, data for some 100,000 galaxies and more than 36,000 redshifts have been collected. From LEDA a volume-limited (v<15,000 km.s$^{-1}$) and diameter-limited (log$D_{25}$ >1.2) all-sky sample of 5,863 galaxies was extracted. After studying its completeness in apparent diameter, we used this sample to show that the most populated plane in the local Universe actually coincides with the 'hypergalactic plane or structure' previously found with the Flamsteed 2D-projection (Di Nella and Paturel, 1994 and 1995). If real, this would represent possibly the largest coherent structure in the Universe.

This plane (hereafter the *hypergalactic* plane) is defined by its updated pole and origin in galactic coordinates ($l(pole) = 52\,\deg; b(pole) = 16\,\deg$ and $l(origin) = l(pole); b(origin) = b(pole) - 90\,\deg$. To understand the significance of this most populated plane we calculated that at $\pm 15\,\deg$ around this plane 45% of the sample lay in only 25% of the solid angle. The distribution of the galaxies of the sample is shown in Fig 1.

Despite the wealth of information in Figure 1 there was still a clear need to improve the picture further via additional redshifts. Like the Fairall redshift compilation, the available LEDA redshifts are a collation from a wide variety of sources all possessing their own selection criteria, biases and errors. Incompleteness in some of these surveys is also occurring just where the walls are becoming populated. Furthermore our azimuthal view is incomplete due to the zones of avoidance in the Milky Way and possible poor sampling in certain directions.

Thus to ascertain whether the great walls seen are part of yet larger shell-like structures we have begun a redshift survey to map the galaxy distribution out through, and beyond the great walls. We hope to determine whether they truly have an outer edge (rather than being an artefact of incomplete data) and also to determine whether the apparent gaps between the walls are real or due to poor sampling. Apart from being pierced by narrow pencil-beam surveys the galaxy distribution beyond the walls has so far been relatively poorly studied.

In this first paper we present results of a new redshift survey in the direction between the Pavo-Indus and Perseus-Pisces walls. This is of particular interest because of the apparent deficiency of galaxies in this direction seen in Figure1. A prime motivator for studying this region first was to see if there is a possible link between these two major walls as suggested by 2D-projections (Paturel et al., 1988; Santiago et al. 1995).

The paper is organised as follows: in section 2 we describe the selection of the galaxies to be observed, the observing procedure and the data reduction. Section 3 presents the final redshift results whilst the significance of our findings are discussed in section 4.

## 2 OBSERVATIONS

### 2.1 Construction of the sample

To obtain our sample, we want to selected galaxies in the empty region between Perseus-Pisces and Pavo-Indus. This region corresponds to the range $150\,\deg \leq hgl \leq 180\,\deg$ and $-10\,\deg \leq hgb \leq +10\,\deg$ in hypergalactic longitude and latitude respectively. We decided to limit ourselves in hypergalactic latitude and study only the galaxies lying close to the hypergalactic plane. This would encompass FLAIR observations across 4 UKST/SERC survey fields centered at approximately hgb=0 deg (see Table 2).

From the LEDA database, which contains published data on known galaxies, we selected all galaxies which are positioned in one of the four UKST fields to be observed. Though entries on more than 100,000 galaxies are currently available from LEDA, significant incompleteness exists for a whole range of LEDA galaxy parameters including redshift, magnitude and apparent diameters. Specially, for galaxies fainter than 15.5 B-magnitude, LEDA becomes progressively incomplete.

Thus, it is necessary to complete the LEDA sample from a deeper catalogue. We used the COSMOS database currently 'on line' at the AAO (Drinkwater, Barnes and Ellison, 1995). The COSMOS database consists of parameterised image information for all objects detected above a given isophotal threshold from COSMOS measuring machine scans of the UKST/SERC J survey plates of the whole of the Southern sky (excluding the Milky Way). Star-galaxy separation is provided up to the limiting magnitude for galaxies of $B_T \sim 21$ though sources can be detected up to $B_T \sim 23$. Howeveer, we only selected galaxies in the range $15.5 \leq B_T \leq 17$. This limit, sufficiently deep to provide good sampling at the depths of interest, is a good match to the capabilities of FLAIR and provides a target list that can be observed in reasonable timescales.

**Table 1.** Details of the observations

| ESO/SERC field # | Set of galaxies | Night of Sept. 1994 | Seeing arcsec. | Total exposure time sec. |
|---|---|---|---|---|
| 346 | a | 6th/7th | 2-4 | 13500 |
| 346 | a | 7th/8th | 1-2 | 9600 |
| 537 | a | 7th/8th | 1-2 | 21000 |
| 290 | a | 8th/9th | 2 | 15000 |
| 470 | a | 8th/9th | 2 | 15000 |
| 346 | b | 9th/10th | 1-2 | 15900 |
| 290 | b | 9th/10th | 1-2 | 7200 |

Some galaxies fainter than $B_T = 15.5$ can be present in both LEDA and COSMOS subsamples (see section 2.5). After matching galaxies in both subsamples and rejectig galaxies with known redshift we obtained a total of 905 galaxies for redshift determination as shown in Table 2. Five nights of FLAIR observing time were allocated for the project in September 1994. Even with the fibre multiplex advantage and the capability to observe $\sim 140$ objects a night by using two FLAIR plateholders it was not possible to observe the total sample of selected galaxies in the nights available. Hence 2 sets of $\sim 80$ galaxies per UKST field were randomly selected at the rate of 1-in-2 as this gives a good match to the available fibres in each FLAIR plateholder. A final sample of 460 galaxies was obtained.

## 2.2 Observing procedure

The survey was undertaken using FLAIR, the multi-fibre spectroscopy system on the 1.2m UK Schmidt telescope at Siding Spring, Coonabarabran, Australia. FLAIR is ideal for this survey as it combines both a large field of view (6 deg $\times$ 6deg), a large multiplex advantage (73-92 objects simultaneously depending on the FLAIR plateholder used) and a sufficiently faint limiting magnitude ($B_T \leq 17$) to efficiently sample the volume of interest.

The observations, spread over five nights from the 5th to the 9th September 1994, are detailed in Table 1. Unfortunately the first night was lost due to poor weather. The second night was only partly clear with typical seeing of 2-5 arcseconds. The following nights were mostly good with seeing in the range1-2 arcseconds. Two sets of 80-85 galaxies per field were prepared with the idea of observing two sets of galaxies per night by swapping between the two available plateholders. This was accomplished on three nights but due to weather conditions and some equipment problems half-night exposures were sometimes not enough to obtain data of adequate S/N. In these cases plateholder changeover was delayed to later in the night or to the following night.

The length of a FLAIR exposure is determined mainly by the cosmic ray event rate, typically 2-3 per minute. After 3000 seconds $\sim 100$ cosmic ray events may be detected resulting in significant clutter of a data frame. Important features in galaxy spectra may thus be compromised. An upper limit of 3000 seconds exposure was adopted. The typical total exposure time per field was in the range $3 \times 3000s$ to $7 \times 3000s$. $5 \times 3000s$ are normally considered adequate to obtain spectra of S/N ratio $\sim$20.

The G300B grating was used giving 232 Å/mm or 5.12 Å/pixel (CCD resolution). The 2850Å covered on the CCD at this dispersion was selected for the range 4400Å – 7360Å. This choice provided:
i) sufficient spectral coverage to give a good chance of seeing a number of different absorption or emission features in galaxies at different redshifts,
ii) a high enough dispersion to get reasonable redshift accuracy,
iii) a minimum spreading of the galaxy light to get an acceptable S/N ratio.

The current FLAIR system has very poor efficiency below 4500 Å, but this problem has recently been addressed by the commissioning of a new back illuminated thinned CCD which offers $> 3$ times the blue DQE of the previous CCD.

To wavelength calibrate the data, arc exposures were taken either side of a set of field exposures at the given dispersion and grating angle. The extremely stable nature of the floor-mounted FLAIR spectrograph obviates the need for more frequent calibration exposures (e.g. Parker & Watson 1995). Light from a selection of arc lamps is simply reflected off the closed dome. Arc exposures were also taken when we changed plateholders to account for the different fibre-formats, focus values etc. Both Neon and Hg-Cd arcs were combined to provide adequate line coverage over the observed wavelength range.

Flat-field exposures were taken using either the zenith twilight sky or by reflection of a featureless quartz-halogen lamp off a specially provided dome flat-field screen. These exposures are used to obtain the fibre-to-fibre transmission function (see Parker and Watson 1990) vital to ensure proper sky-subtraction. Differences between fibre transmission efficiencies between the twilight sky and dome flats are at the $\sim$ 2% level (Parker & Lee, 1994). Bias frames were also taken before and/or after observations in order to correct for any sensitivity variations, cosmetic defects and systematic pixel-to-pixel variations on the CCD. Of the 92 or 72 available fibres (depending on which FLAIR plateholder is used), 5-8 were devoted to the night sky to facilitate satisfactory sky subtraction across the wide field.

## 2.3 Data reduction

All the data reduction was performed using the NOAO IRAF spectral reduction package together with a few additional FLAIR specific IRAF tasks. A FLAIR IRAF data reduction manual exists to facilitate the process (Drinkwater and Barnes, 1994). It is based mainly on existing IRAF packages for multifiber spectroscopy such as 'dohydra'. The final derivation of the radial velocities was done using the 'rvsao' cross-correlation IRAF package.

The first stage of the process is to combine several bias exposures into a single image; this is required to remove any structure in the bias level across the CCD. It is also necessary to combine the flat field frames used to determine and correct for the relative transmission efficiency of the different fibre apertures. This frame is also used to define and identify the fibre apertures themselves. The next stage is to use the IRAF task 'ccdproc'. This task removes the electronic zero level first by subtracting the combined bias image from each frame and secondly by subtracting the average over the columns in the overscan region. Finally the image is trimmed to leave just the part containing useful data. The individual data frames are now combined to increase the S/N ratio of the final spectra. At this stage the cosmic ray events on each frame are effectively removed using a suitable sigma clipping algorithm which rejects pixel whose values deviate significantly from the mean with minimal loss of data. An aperture identification file is then created to specify which apertures are sky and which contain the objects. The FLAIR data are now ready for final processing using the IRAF 'dohydra' task. This complex task was developed for the HYDRA multi-fibre system on the KPNO 4m telescope but can also be effectively used with FLAIR data. From the combined, cleaned and trimmed target frames together with the dome flats and arc exposures, fully reduced dispersion corrected (wavelength calibrated), sky subtracted one dimensional spectra were produced.

**Table 2.** Selection of the target galaxies

| HGB range in deg | HGL range in deg | RA and DEC of field centre | UKST/SERC field number | number of target objects | number of observed objects |
|---|---|---|---|---|---|
| -6 +0 | 174 180 | 23h50 -25deg | 537 | 199 | 83 |
| -3 +3 | 166 172 | 23h23 -30deg | 470 | 202 | 85 |
| -3 +3 | 154 160 | 22h58 -40deg | 346 | 247 | 129 |
| -5 +1 | 149 155 | 22h52 -45deg | 290 | 257 | 163 |

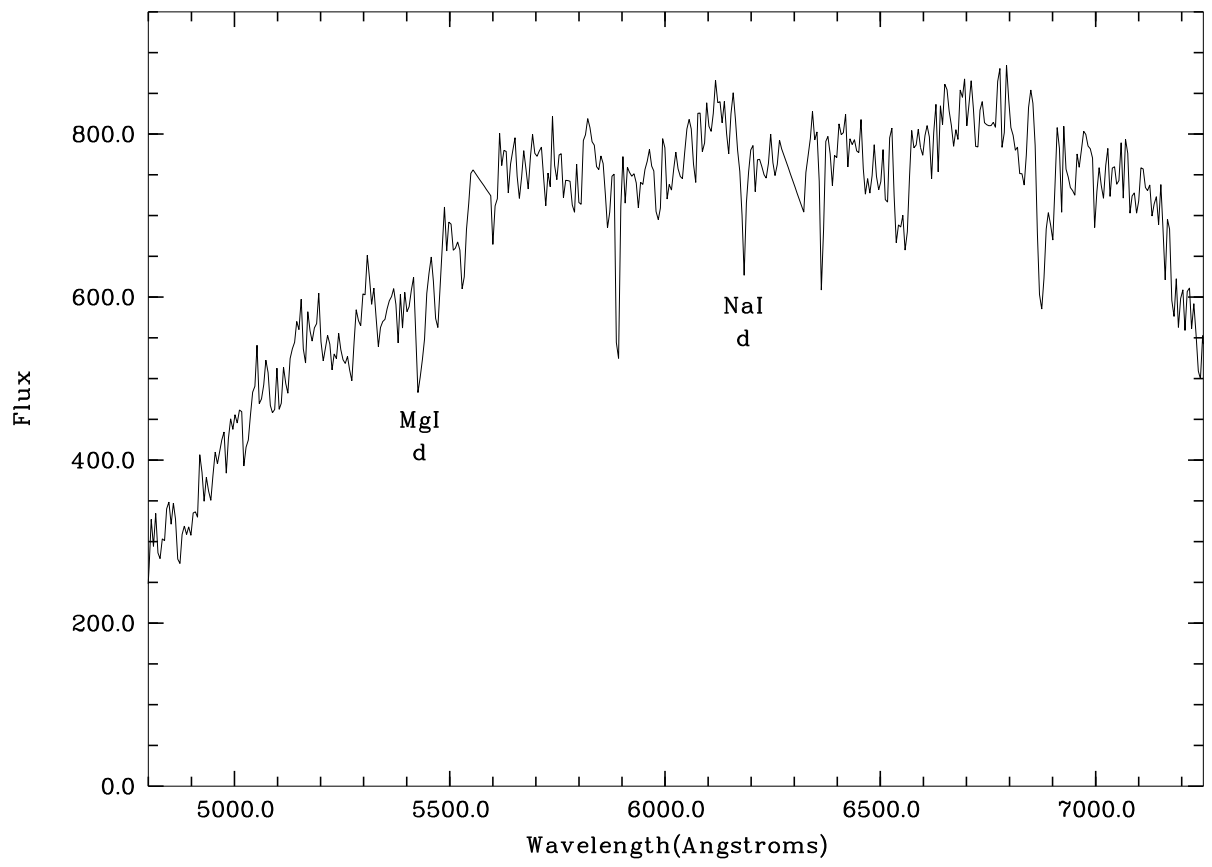

**Figure 2.** Typical reduced absorption line galaxy spectrum (not flux- calibrated) with S/N ratio ∼18.5. The vertical axis average counts per exposure ($1e^{-} = 1edu$). The apparent magnitude is $B_T = 15.3$. The combined exposure was $7 \times 3000s$ and $V_{helio} = 14772 km.s^{-1}$.

## 2.4 Derivation of radial velocities

The reduced spectra from 'dohydra' can then be input to the IRAF 'rvsao' task to derive the galaxy radial velocities via cross-correlation against filtered templates provided by Q.A. Parker. Such reduction procedures are described in detail by Parker and Watson (1990) and Watson et al. (1991,1992).

We finally observed 6 sets of ∼ 80 galaxies (fields 346 and 290 were observed twice but looking at different galaxies). Among the 460 galaxies observed, 379 gave reliable radial velocities according to a set of strict criteria: ccf peak height cut-off and match between emission and absorption line velocities). An overall success rate of 82% was achieved.

Two fully reduced typical spectra are shown in Figure 2 and Figure 3 for an absorption and emission line spectrum respectively. During the survey a significant number of active galaxies were also identified; they will be reported in a companion paper.

The list of the 379 measured galaxies is given in Table 3 together with COSMOS coordinates and apparent magnitude. These parameters (plus diameter, axis ratio and position angle) were used to perform cross-identification with the PGC/LEDA objects after reduction to the RC3 system (de Vaucouleurs et al., 1990). Both cross correlation results (from 'rvsao') and emission line results (from 'emsao') are presented. A few galaxies with published redshifts were also observed to derive our external error (see section 3).

The columns of Table 3 are arranged as follows:

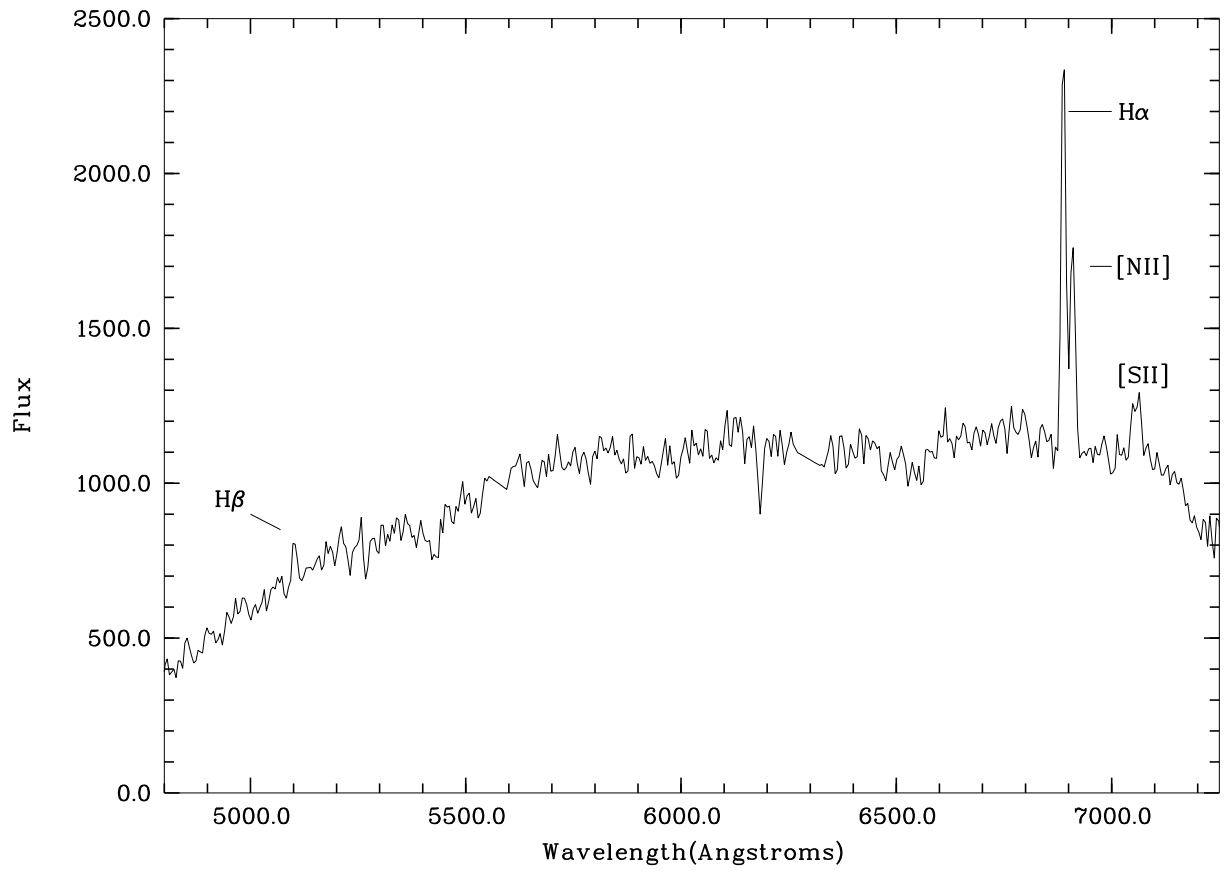

**Figure 3.** Typical reduced emission line galaxy spectrum (not flux calibrated) with Hα, $H_\beta$ and S[II] lines.The S/N ratio is ∼21. The vertical axis is average counts per exposure. The apparent magnitude is $B_T = 12.6$. The combined exposure is $7 \times 3000s$ and $V_{helio} = 14839 km.s^{-1}$.

- **column 1:** Right ascension in hours, minutes, seconds, tenths for Equinox 1950.0.
- **column 2:** Declination in degrees, arcminutes and arcseconds for Equinox 1950.0. Coordinates are from the COSMOS database except for a few cases were positions were taken from the LEDA database.
- **column 3:** Internal field number

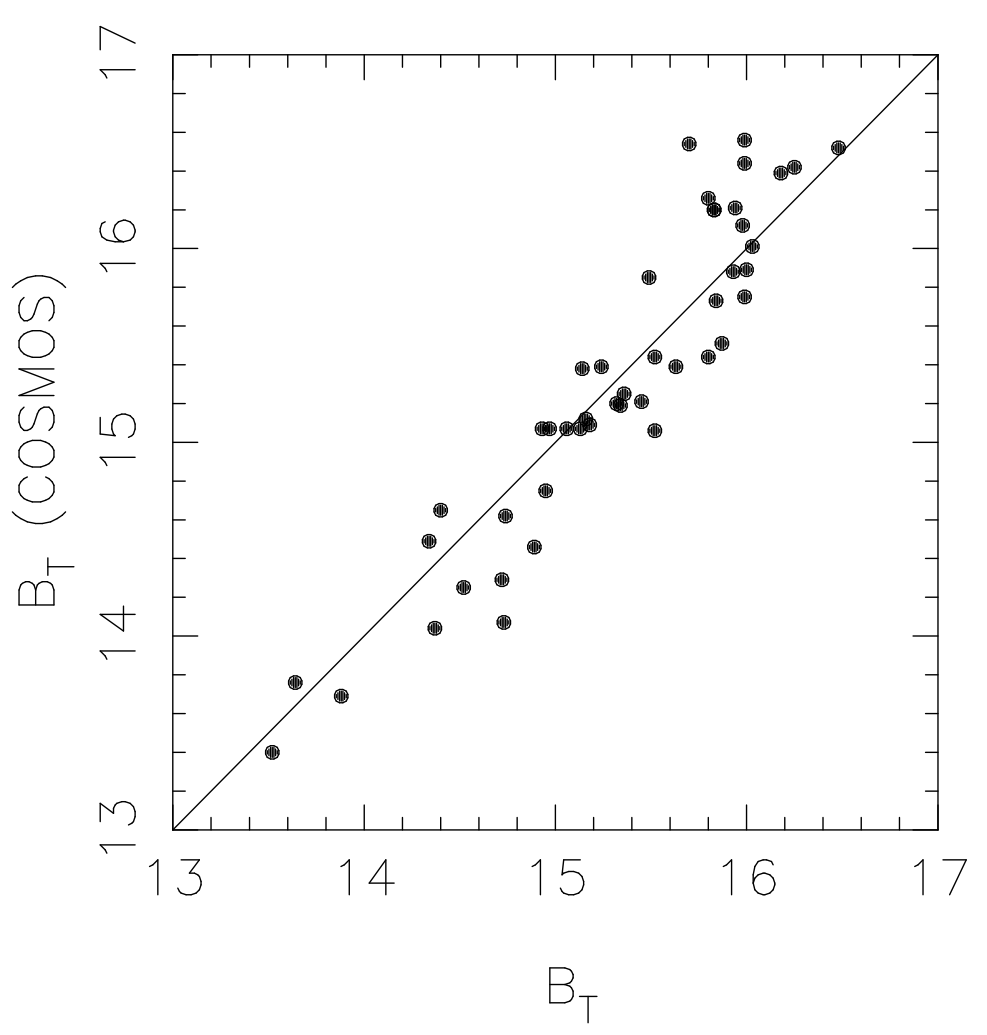

**Figure 4.** Comparison of galaxy B-magnitudes (see text)

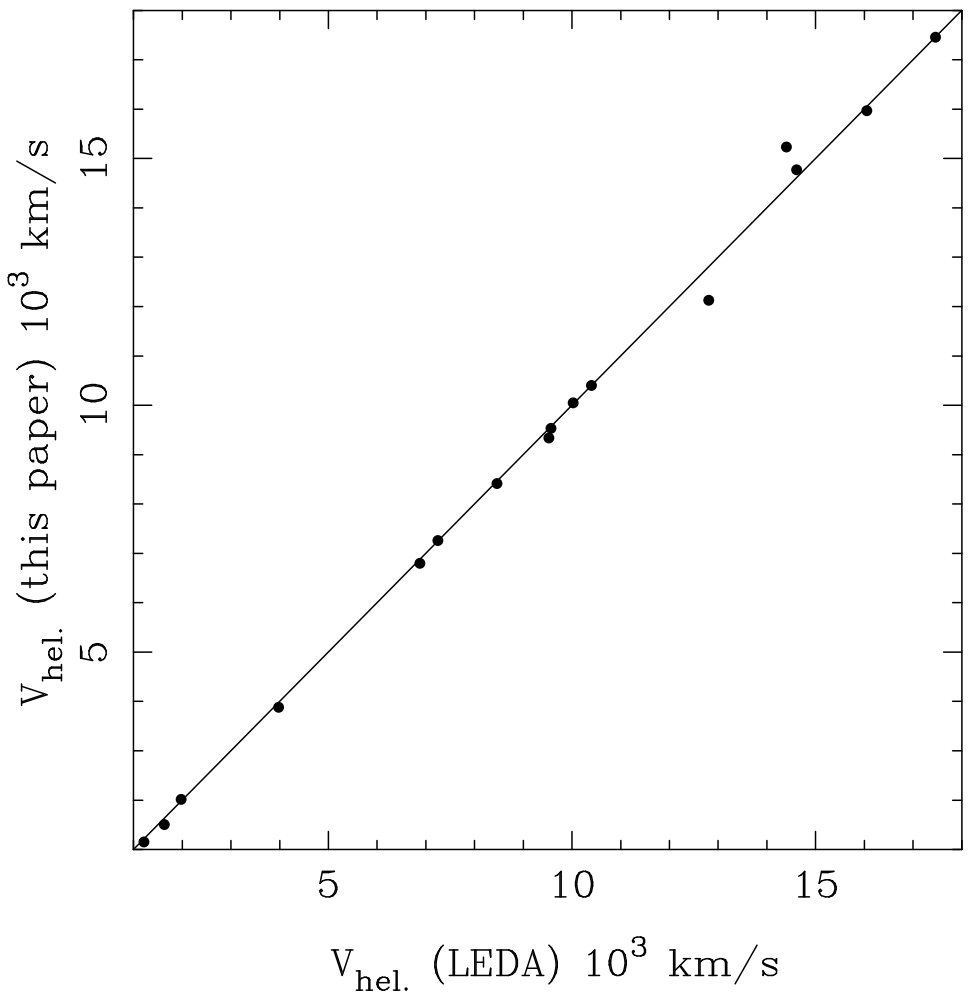

**Figure 5.** Comparison of new and published heliocentric velocities (see text)

- **column 4:** PGC/LEDA number (Paturel et al. 1989a, b).
- **column 5:** Alternative name in a given hierarchy according to LEDA database: NGC, IC, ESO, MCG, FAIR and DRCG (Dreyer, 1889-1910; Lauberts 1973-1982; Vorontsov-Velyaminov et al. 1962-1974; Fairall 1988, and Dressler 1980).
- **column 6:** $B_T$ magnitude from COSMOS database after reduction to the LEDA magnitude system (Figure 4): $B_T = B_{COSMOS} - 0.13$
- **column 7:** Heliocentric velocity $V_{cc}$ from the cross-correlation method, and its mean error (in $km.s^{-1}$).
- **column 8:** Heliocentric velocity $V_{em}$ from the emission lines identification method, and its mean error (in $km.s^{-1}$)
- **column 9:** notes

### 2.5 Radial velocity external error estimates

To determine the external mean error we made a comparison (Fig 5) between our new measured velocities and those available in LEDA for the few objects in common. This comparison confirms that there is no significant error for the slope or for the zero-point. Rejecting two uncertain measurements, the residual error is $\sigma$=84 km.s$^{-1}$. If we assume that both LEDA and the measurements of this survey have the same mean error, we get a mean error of $\sigma(V_{cc})$=59 km.s$^{-1}$.

The heliocentric velocities measured by the emission line detection method are also compared to published values despite there being only 5 points in common. No slope or zero-point error is found. This is confirmed by a comparison with radial velocities obtained by the cross-correlation method. The mean error is tentatively $\sigma \approx 180 km.s^{-1}$.

## 3 RESULTS

The aim of this survey was to study the distribution of galaxies in the region between Perseus-Pisces and Pavo-Indus (see figure 1). With this study we hoped to ascertain whether the apparent gap between these two walls seen in Figure 1 is:

- due to poor sampling of this region by previous surveys, and therefore if there is a link between the Perseus-Pisces and Pavo-Indus Superclusters.
- due to a real void of galaxies.
- due to an 'apparent' void of galaxies caused by high-latitude galactic absorption.

To address these questions we first present the new results of this survey in Figure 10. It is important to understand that the plane of this cone diagram is exactly the same as the plane of Figure 1 (the 'hypergalactic' or most populated plane). There is evidence of two density enhancements at a radial distance of $\sim$ 133 and $\sim$ 200 Mpc. Their significance seem more clear with the help of a redshift histogram shown in Figure 7. It appears in Figure 11 that these enhancements delineate two low density regions or 'voids'. When compared to the maps of this region given in the compilation of Fairall & Jones (1991) these voids are real and significant. The nearest one with v< 10,000 km.s$^{-1}$ is known as the Sculptor Void and the furthest with v> 10,000 km.s$^{-1}$ was called the Further Sculptor Void by Fairall & Jones. This figure is very interesting because it represents one of the rare regions where two walls perpendicular to the line of sight are seen. We call these 'chains' seen from our survey 'walls' because they are part of structures which extend above and under the plane of this cone diagram (see Fairall's catalogue of maps of this region).

When we add these new redshifts to the data of Figure1, no evidence of a link between the northern Perseus-Pisces and the southern Pavo-Indus features can be seen. The current survey sampled at a rate of 1-in-2 is highly significant and thus we conclude that we cannot have 'missed' any link if one existed.

The most prominent feature in Figure 11 is the double peaked velocity distribution. The first peak at $V_{hel}$ $\approx$10,000 km.s$^{-1}$ seems to result from galaxies located in a southern extension of the Perseus Pisces wall which now extends over more than 150 Mpc. The second peak is located at $V_{hel}$ $\approx$ 15,000 km.s$^{-1}$.

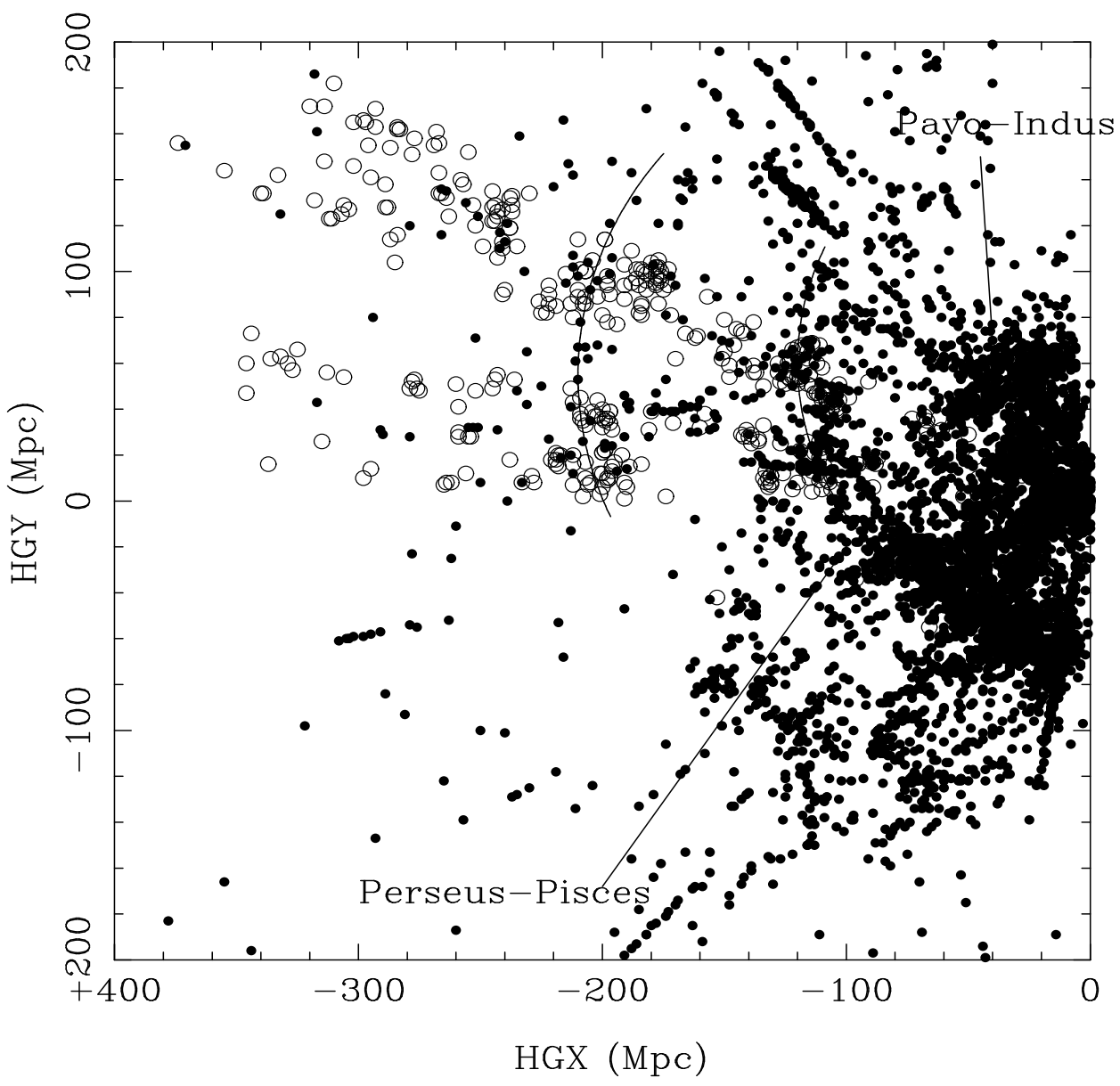

**Figure 6.** Distribution of galaxies within the region surveyed in this study. The filled circles represent LEDA galaxies previously measured in redshift, the open circles are from this survey. This is a view face-on of the hypergalactic plane plane with the Pavo-Indus wall and Perseus-Pisces chain. The position of the density peaks seen in the velocity histogram are indicated by the circle arcs. One of this velocity peak in the redshift survey shows a southern extension to Perseus-Pisces which now extends over more than 150 Mpc.

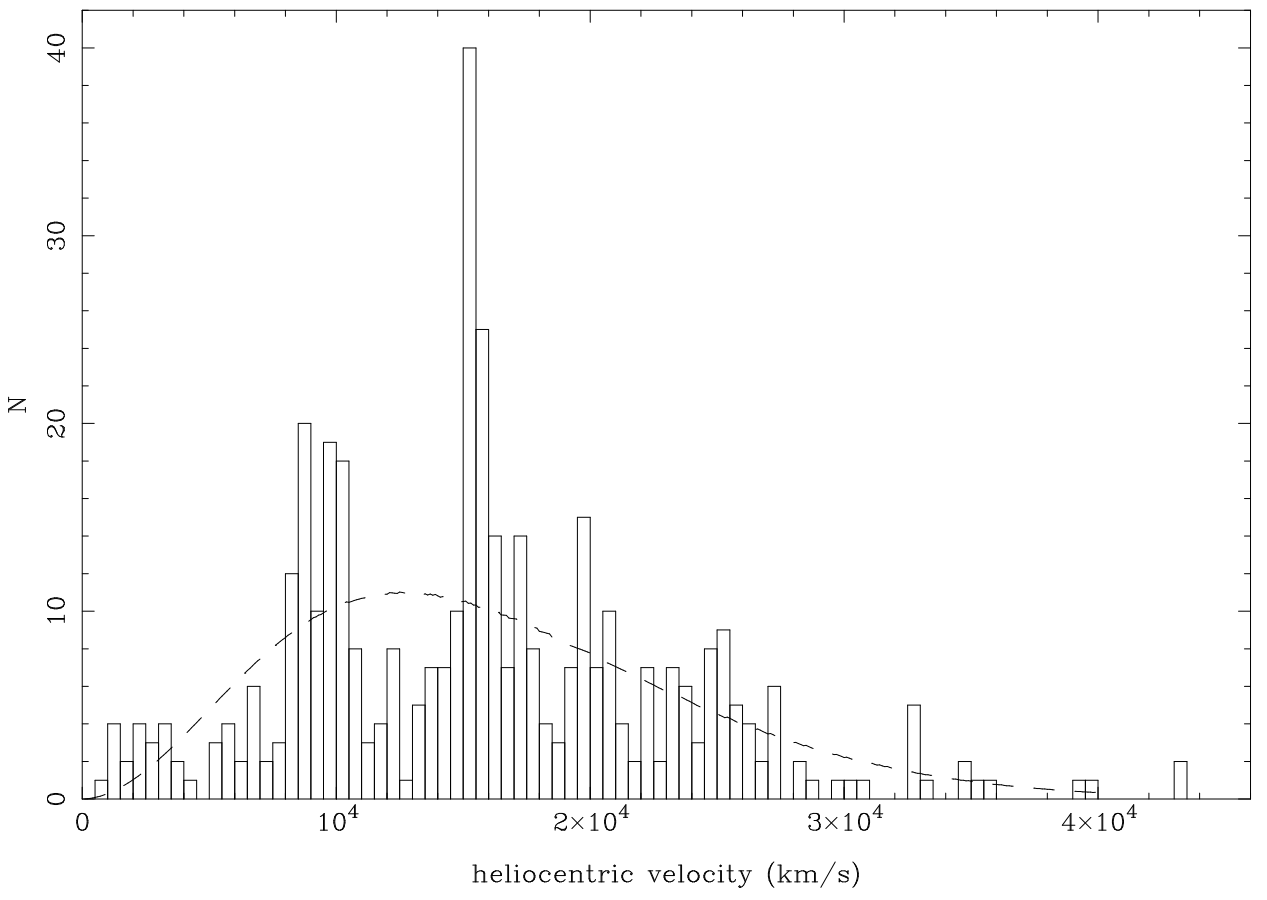

**Figure 7.** Histogram of heliocentric velocities in the studied region. The dashed line represents the selection function of our survey. The most important features are the two highly significant peaks bracketting under-populated regions.

They are compared to the calculated selection function of our survey derived from a standard non- evolving galaxy luminosity function. Because we are claiming that our sample is complete between $B_T = 15.5$ and $B_T = 17$, we calculated the selection function between these limits, using a Schechter function (Schechter and Press, 1976) with the standard parameters $M_* = -20$ and $\alpha = -1.07$. The function is normalized to our sample total of $\approx 400$ galaxies.

Assuming Poisson errors, both peaks are significant at a level of $5\sigma$ and $11\sigma$, respectively. It is also clear that the regions corresponding to the Sculptor and the Further Sculptor voids are not consistent with a uniform average space density of galaxies but really are under-populated regions.

## 4 DISCUSSION AND CONCLUSIONS

One could argue that the observed galaxy distribution is the result of selection effects. Indeed, if our sample was chosen in a biased way, e.g. with more time devoted to a given region at a given distance, one could suspect that the peaks simply reflect the excess of time spent at this distance. Several arguments can be presented against this interpretation.

- Our selection was made on the basis of carefully controlled criteria, $15.5 \leq B_T \leq 17$, and without regard to the local density or to distance. Thus, the observed density at a given distance should be representative. In interpreting features of the redshift distribution in this way, one must be extremely worry about the selection effects which bias our view of the structure that is present. For example, could the peaks observed in the velocity distribution simply indicate that our sample is preferentially "'tuned" to galaxies at these distances? From the selection function we have computed, the answer in our case is no.
- A completeness test can be made to prove that our sample constitutes a magnitude limited sample with a well defined limit (Figure 8). This diagram has the expected slope of 0.6 indicating that the number of studied galaxies increases, on the mean, as the cube of the distance. For comparison, we plotted on the same figure the completeness curve for the entire sky (normalised to the solid angle of the present survey) obtained from LEDA when imposing that both the apparent magnitude $B_T$ and the radial velocity $v$ are known. The completeness limit in the studied region is 1.5 magnitude deeper after our survey.
- Another argument comes from the diagram Figure 6 which shows that the density enhancements are visible in each covered field at the same distance.
- Finally the nearest of the two walls was already visible in the cone diagrams compiled by Fairall and Jones (1991). This wall delineates the Sculptor void and the Further Sculptor void. It is perpendicular to the Sculptor Wall. The furthest wall (at 15,000 km.s$^{-1}$) has never been noticed before. It delineates the end of the Further Sculptor Void and is perpendicular to the Scuptor Wall.

To continue to address the questions exhibited in section 4 we have explored the map of Burstein & Heiles (1984) to see if a high galactic latitude molecular cloud could be responsible for an apparent void in the distribution of galaxies such as the Sculptor Void. If this was the case, the possibility of a hidden connection between Perseus-Pisces and Pavo-indus could still exist. However we have already a hint: Fairall & Jones didn't see a second wall delimiting the end of the Further Sculptor void, but we did. So it would be very surprising if the galactic absorption could play a role depending on distance i.e. hiding some near galaxies and not some more distant ones. Reinforcing this answer, no mapped

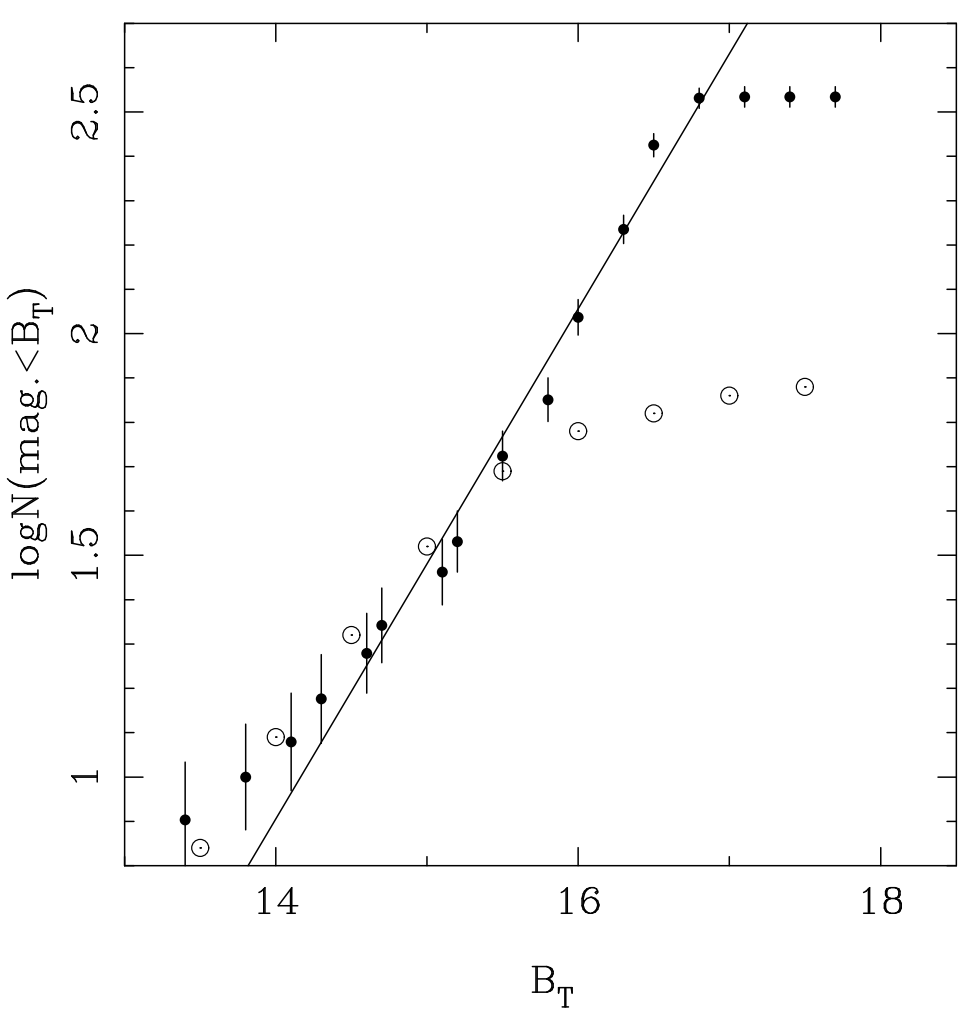

**Figure 8.** Test to show that our sample constitutes a magnitude limited sample with a well-defined cut-off at $B_T = 17$. The solid line represents the expected slope 0.6. The mean completeness curve obtainied from LEDA for the same solid angle is shown in open circles. A gain of 1.5 magnitude results from our survey.

absorbing cloud is present in the available absorption maps in this direction.

In conclusion, we no longer suspect a link between the Perseus- Pisces and Pavo-Indus walls despite evidences from 2D-projections (Paturel et al., 1988; Santiago et al. 1995). Nevertheless Figure 1 is still a very good representation of our neighbourhood. Updating this figure will be done only by studying the outer regions of the major walls and trying to fill in the missing data across the zones of avoidance. The Further Sculptor Void can now be considered delimited by a wall at 15,000 km.s$^{-1}$, and Perseus-Pisces is extended in the Southern hemisphere. The total length of this chain is now more than 150 Mpc.

## ACKNOWLEDGEMENTS

We are grateful to those who have managed the LEDA extragalactic database over the last 12 years: N. Durand, P. Fouqué, A.M. Garcia, R.Garnier, M. Loulergue, M.C. Marthinet and C. Petit. We are also grateful to those who assisted during the observations: the UKST staff, S. Besier and K. Sealey. We acknowledge the COSMOS/UKST Southern Sky Catalogue supplied by the COSMOS group at the Royal Observatory Edinburgh to the Anglo-Australian Observatory. This work has been supported by grants from the Conseil Régional Rhone-Alpes (F) and by the Institut National des Sciences de l'Univers (F). We thank the University of New South Wales for welcoming us and for providing us the facilities with which we have conducted this study. Q.A. Parker is on special leave from the ROE to work at the AAO. We thank the referee for his comments.

Table 3: The 379 measured galaxies.

| RA 1950 DEC<br>h m s d ' "<br>(1) (2) | field number<br>(3) | PGC/LEDA number<br>(4) | Alternat. name<br>(5) | BT<br>(6) | Vcc km/s<br>(7) | Err km/s | Vem km/s<br>(8) | Err km/s | Notes<br>(9) |
|---|---|---|---|---|---|---|---|---|---|
| 223534.39-453312.0 | 290a_1 | | | 15.52 | 15343 | 93 | - | - | abs |
| 223606.60-422352.2 | 290a_2 | PGC069407 | ESO345-35 | 16.39 | 17211 | 48 | - | - | abs |
| 223617.59-422803.7 | 290b_104 | | | 16.58 | 12022 | 59 | 12008 | 53 | emi |
| 223625.52-455528.6 | 290b_105 | | | 16.70 | 19921 | 146 | 19920 | 54 | emi |
| 223638.12-453403.8 | 290b_106 | | | 16.67 | 7881 | 67 | - | - | abs |
| 223647.90-453951.8 | 290a_8 | | | 16.53 | 15576 | 37 | 15574 | 27 | emi |
| 223651.77-455210.1 | 290a_7 | | | 16.39 | 10236 | 28 | 10401 | 22 | emi |
| 223803.00-473548.0 | 290b_108 | PGC069476 | ESO238-21 | 17.35 | 9726 | 50 | - | - | emi Sb |
| 223812.13-470207.8 | 290a_11 | | | 16.03 | 9998 | 56 | 10146 | 8 | emi |
| 223824.69-444919.8 | 290b_109 | | | 16.65 | 15066 | 81 | - | - | abs |
| 223825.80-452727.1 | 290a_12 | | | 16.30 | 23925 | 57 | 23941 | 56 | emi |
| 223906.86-443008.5 | 290a_14 | | | 16.20 | 15057 | 67 | 15274 | 9 | abs+emi |
| 223943.49-431014.0 | 290b_211 | PGC069554 | ESO290-4 | 14.62 | 9339 | 65 | - | - | abs S0 9522 |
| 224013.48-465357.0 | 290a_15 | | | 16.06 | 9967 | 46 | 10024 | 10 | abs+emi |
| 224029.38-464004.2 | 290a_16 | | | 16.33 | 27004 | 81 | - | - | abs |
| 224036.80-453523.4 | 290b_212 | PGC069589 | ESO290-6 | 13.82 | 2883 | 54 | 2866 | 16 | emi Scd |
| 224039.16-440930.7 | 290b_113 | PGC069593 | ESO290-7 | 15.20 | 9531 | 46 | 9950 | 28 | emi Sb 9568 |
| 224126.66-445833.4 | 290a_17 | | | 16.52 | 20948 | 55 | 21024 | 13 | emi |
| 224131.61-454921.9 | 290b_116 | LEDA100667 | | 16.77 | 13531 | 50 | - | - | abs |
| 224209.03-471143.8 | 290b_117 | | | 16.86 | 15440 | 72 | 15548 | 69 | emi |
| 224215.45-473752.0 | 290b_118 | | | 16.96 | 23453 | 101 | 23422 | 31 | abs+emi |
| 224244.22-411047.1 | 346a_1 | PGC069665 | ESO345-50 | 15.34 | 1772 | 77 | - | - | abs |
| 224329.89-462214.1 | 290a_19 | | | 16.30 | 15438 | 72 | 15317 | 55 | emi |
| 224345.51-440042.2 | 290b_120 | PGC069698 | ESO290-11A | 16.65 | 20677 | 74 | - | - | abs |
| 224404.32-434556.6 | 290a_20 | | | 16.17 | 20700 | 136 | 20717 | 25 | abs+emi |
| 224406.63-460803.3 | 290b_122 | | | 16.79 | 15162 | 74 | - | - | abs |
| 224422.09-465132.3 | 290a_21 | | | 16.37 | 16284 | 43 | 16310 | 7 | abs+emi |
| 224423.80-380845.2 | 346b_3 | | | 17.00 | 9218 | 154 | 9305 | 29 | emi |
| 224427.63-453347.7 | 290b_123 | | | 16.83 | 20426 | 119 | 20672 | 39 | abs+emi |
| 224428.59-415016.2 | 346b_4 | | | 16.07 | 17438 | 40 | 17703 | 25 | emi |
| 224437.73-462353.2 | 290b_224 | | | 16.88 | 24537 | 64 | - | - | abs |
| 224510.13-414032.7 | 346a_7 | | | 16.97 | 24008 | 109 | 23838 | 100 | emi |
| 224513.96-411603.0 | 346b_7 | | | 16.96 | 21785 | 293 | 21945 | 13 | emi |
| 224526.96-450637.1 | 290b_228 | | | 16.86 | 15545 | 81 | - | - | abs |
| 224540.58-392646.2 | 346b_93 | | | 16.99 | 2734 | 36 | 2790 | 18 | emi |
| 224544.57-460838.8 | 290b_129 | | | 16.47 | 11855 | 23 | 12000 | 16 | emi |
| 224553.72-373323.6 | 346b_10 | | | 16.23 | 8877 | 57 | 8635 | 45 | emi |
| 224603.91-430511.8 | 290b_131 | PGC069776 | ESO290-15 | 16.67 | 5112 | 20 | 5161 | 17 | emi |
| 224612.87-444348.6 | 290a_26 | | | 16.24 | 24390 | 114 | - | - | abs |
| 224613.53-373751.7 | 346b_12 | | | 15.90 | 8656 | 25 | 9045 | 10 | emi |
| 224613.88-385121.2 | 346b_92 | | | 16.71 | 9099 | 26 | 9177 | 6 | emi |
| 224615.42-463623.6 | 290b_132 | | | 16.72 | 24434 | 95 | - | - | abs |
| 224617.76-401448.3 | 346b_96 | | | 16.49 | 9763 | 25 | 9921 | 13 | emi |
| 224629.95-451224.9 | 290b_232 | | | 16.74 | 17916 | 166 | 17583 | 32 | emi |
| 224636.90-374101.5 | 346a_13 | | | 16.30 | 8740 | 76 | - | - | abs |
| 224648.20-471413.1 | 290a_27 | | | 16.58 | 23172 | 55 | - | - | abs |
| 224649.17-385745.2 | 346a_14 | | | 16.55 | 11265 | 95 | 11326 | 21 | emi |
| 224707.60-374847.9 | 346b_14 | | | 16.63 | 8785 | 104 | - | - | abs |
| 224710.25-383515.5 | 346a_15 | | | 16.37 | 24750 | 100 | - | - | abs |
| 224710.54-414229.0 | 346b_70 | | | 16.75 | 9562 | 45 | 9652 | 3 | emi |
| 224719.62-443529.8 | 290a_30 | | | 16.20 | 9876 | 22 | 9911 | 23 | emi |
| 224722.83-465029.7 | 290b_135 | | | 16.71 | 24601 | 105 | 24368 | 51 | emi |
| 224722.95-430254.7 | 290a_31 | PGC069833 | ESO290-17 | 15.57 | 19980 | 71 | - | - | abs S0 |
| 224734.23-440435.8 | 290a_32 | PGC069843 | ESO290-18 | 15.86 | 20539 | 134 | - | - | abs+emi |
| 224735.79-374834.6 | 346a_16 | | | 15.84 | 8689 | 82 | - | - | abs |

Table 3: (continued)

| RA 1950 DEC<br>h m s d ' "<br>(1) (2) | field<br>number<br>(3) | PGC/LEDA<br>number<br>(4) | Alternat.<br>name<br>(5) | BT<br><br>(6) | Vcc<br>km/s<br>(7) | Err<br>km/s | Vem<br>km/s<br>(8) | Err<br>km/s | Notes<br><br>(9) |
|---|---|---|---|---|---|---|---|---|---|
| 224749.02-453546.5 | 290b_137 | PGC069849 | ESO290-20 | 14.17 | 15226 | 77 | - | - | abs E |
| 224806.59-403117.8 | 346b_16 | | | 15.64 | 10049 | 52 | 10037 | 16 | emi |
| 224811.90-453400.3 | 290b_238 | | | 16.86 | 26880 | 46 | - | - | abs |
| 224814.78-465618.0 | 290a_33 | | | 15.65 | 23289 | 86 | - | - | abs |
| 224816.38-462421.1 | 290a_34 | | | 16.03 | 10141 | 78 | - | - | abs+emi |
| 224822.88-380538.4 | 346a_18 | | | 16.63 | 24863 | 149 | 24989 | 147 | emi |
| 224823.07-382249.6 | 346b_18 | PGC069864 | ESO346-5 | 16.02 | 9047 | 52 | 9062 | 79 | emi Sa |
| 224857.73-395214.6 | 346b_19 | | | 16.03 | 17218 | 36 | 17233 | 18 | emi |
| 224903.80-403351.7 | 346b_82 | | | 16.86 | 9803 | 40 | 9955 | 23 | emi |
| 224915.89-453839.0 | 290a_38 | | | 16.49 | 15756 | 87 | - | - | abs |
| 224932.92-453401.0 | 290b_141 | | | 16.87 | 23761 | 69 | 23875 | 5 | emi |
| 224932.98-403440.6 | 346b_81 | | | 15.87 | 9827 | 45 | 9896 | 15 | emi |
| 224937.58-381412.0 | 346b_20 | | | 16.59 | 8441 | 64 | - | - | abs |
| 224951.23-454130.7 | 290a_39 | | | 16.29 | 15128 | 59 | - | - | abs |
| 225001.95-373652.0 | 346a_21 | | | 16.40 | 1479 | 43 | 1599 | 41 | emi |
| 225006.34-411148.1 | 346b_69 | | | 16.54 | 16234 | 62 | 16410 | 3 | emi |
| 225007.32-453844.5 | 290a_40 | | | 16.18 | 20238 | 112 | - | - | abs |
| 225026.04-394632.3 | 346b_21 | | | 16.99 | 16880 | 102 | - | - | abs |
| 225034.21-390356.7 | 346b_94 | PGC069923 | ESO346-7 | 15.20 | 863 | 35 | 928 | 17 | emi Irr |
| 225041.27-374722.1 | 346a_21 | | | 17.00 | 23199 | 54 | - | - | abs |
| 225050.77-440031.1 | 290b_142 | | | 16.77 | 21305 | 68 | 21473 | 30 | emi |
| 225054.30-383359.4 | 346b_22 | | | 16.93 | 12215 | 102 | 12230 | 16 | abs |
| 225105.04-474046.7 | 290b_143 | | | 16.85 | 25449 | 187 | 24754 | 100 | emi |
| 225108.02-391408.8 | 346a_23 | | | 16.21 | 24890 | 50 | - | - | abs |
| 225124.23-402455.8 | 346a_24 | | | 16.07 | 9470 | 9 | 9596 | 15 | emi |
| 225127.88-444427.5 | 290a_41 | | | 15.98 | 25034 | 64 | 25184 | 29 | emi |
| 225128.79-393453.7 | 346b_83 | PGC069964 | NGC7404 | 13.89 | 2017 | 74 | - | - | absE-So 1964 |
| 225130.15-440356.8 | 290a_42 | | | 16.37 | 20507 | 83 | - | - | abs |
| 225140.67-450548.6 | 290a_43 | | | 16.33 | 15126 | 38 | 15071 | 25 | emi |
| 225141.37-423138.9 | 346a_25 | PGC069973 | ESO290-23 | 16.25 | 17258 | 56 | 17541 | 23 | emi |
| 225141.63-473426.7 | 290b_244 | PGC069975 | ESO239-3 | 16.62 | 15054 | 124 | - | - | abs |
| 225210.79-400906.3 | 346b_25 | | | 15.52 | 13280 | 48 | - | - | abs |
| 225233.89-381821.5 | 346b_26 | LEDA70005 | | 14.50 | 11937 | 167 | 12227 | 234 | emi |
| 225247.27-405519.9 | 346a_27 | | | 15.75 | 13200 | 40 | 13344 | 44 | emi |
| 225254.55-425514.5 | 290b_146 | | | 16.71 | 14759 | 49 | - | - | abs |
| 225301.44-450634.3 | 290a_47 | | | 16.55 | 20432 | 51 | - | - | abs+emi |
| 225305.55-450154.7 | 290a_48 | | | 16.12 | 14837 | 52 | - | - | emi |
| 225310.60-421803.7 | 346a_28 | | | 16.53 | 16449 | 109 | - | - | abs |
| 225338.05-393839.3 | 346a_30 | LEDA89332 | | 16.82 | 8412 | 20 | 8469 | 9 | emi |
| 225405.54-440139.8 | 290b_147 | PGC070085 | IC5267B | 13.53 | 1502 | 90 | - | - | abs 1664 |
| 225408.67-451236.3 | 290a_49 | | | 16.31 | 12694 | 131 | - | - | emi |
| 225411.75-425845.9 | 290b_247 | | | 16.82 | 25206 | 50 | 25211 | 16 | emi |
| 225416.91-465516.7 | 290a_50 | | | 16.18 | 20431 | 93 | - | - | abs+emi |
| 225433.12-452618.6 | 290a_51 | | | 16.13 | 20119 | 26 | 20226 | 13 | emi |
| 225501.18-464709.2 | 290a_52 | | | 16.17 | 25602 | 90 | 25619 | 119 | emi |
| 225505.34-450411.4 | 290b_149 | PGC070120 | ESO290-30 | 15.20 | 15738 | 38 | 15701 | 27 | emi |
| 225508.77-374704.7 | 346a_32 | PGC070119 | ESO346-20 | 16.33 | 8358 | 12 | 8425 | 16 | emi |
| 225528.17-391255.8 | 346b_88 | PGC070138 | ESO346-21 | 15.22 | 8625 | 54 | 8718 | 29 | emi |
| 225528.27-461512.5 | 290a_53 | | | 16.36 | 15631 | 76 | 16093 | 44 | abs+emi |
| 225537.54-441931.5 | 290a_55 | | | 16.35 | 15330 | 114 | - | - | abs |
| 225538.50-403322.8 | 346b_76 | | | 16.46 | 10146 | 74 | - | - | abs |
| 225542.03-475630.5 | 290b_150 | PGC070155 | ESO239-5 | 14.73 | 5404 | 38 | 5564 | 1 | emi |
| 225544.66-454525.2 | 290a_56 | | | 16.00 | 21925 | 64 | - | - | abs |
| 225554.22-452420.6 | 290b_250 | | | 16.32 | 15407 | 131 | - | - | emi |
| 225557.37-392554.4 | 346a_33 | | | 16.62 | 10107 | 43 | 10227 | 30 | emi |

Table 3: (continued)

| RA 1950 DEC | field | PGC/LEDA | Alternat. | BT | Vcc | Err | Vem | Err | Notes |
|---|---|---|---|---|---|---|---|---|---|
| h m s d ' " | number | number | name | | km/s | km/s | km/s | km/s | |
| (1) (2) | (3) | (4) | (5) | (6) | (7) | | (8) | | (9) |
| 225610.93-453932.3 | 290a_57 | | | 16.62 | 15143 | 62 | - | - | abs |
| 225615.99-391652.3 | 346b_89 | | | 16.87 | 10308 | 90 | 10394 | 33 | emi |
| 225623.25-461744.2 | 290a_58 | | | 16.29 | 14764 | 70 | 14604 | 11 | emi |
| 225638.95-465420.9 | 290a_59 | | | 16.02 | 10005 | 127 | - | - | abs+emi |
| 225756.39-373630.9 | 346a_38 | | | 16.11 | 8361 | 69 | 8401 | 29 | emi |
| 225818.70-462435.3 | 290b_156 | | | 16.72 | 15877 | 131 | 15940 | 80 | emi |
| 225826.79-402130.4 | 346a_39 | | | 16.88 | 18063 | 76 | - | - | abs |
| 225852.98-412404.0 | 346b_39 | | | 16.75 | 13595 | 13 | 13919 | 16 | emi |
| 225902.48-470619.8 | 290b_157 | | | 16.62 | 15466 | 81 | - | - | abs |
| 225903.18-443518.2 | 290b_257 | | | 16.57 | 15594 | 83 | - | - | abs |
| 225907.41-405535.0 | 346b_72 | PGC070293 | ESO346-24 | 10.37 | 10046 | 31 | 10196 | 1 | emi |
| 225907.43-444231.7 | 290a_60 | | | 16.31 | 15819 | 47 | 15729 | 24 | emi |
| 225923.27-471617.5 | 290a_61 | | | 15.79 | 10336 | 32 | 10416 | 16 | emi |
| 225923.89-394919.3 | 346a_42 | LEDA70304 | | 16.34 | 1155 | 50 | 1218 | 30 | emi |
| 225925.67-384736.0 | 346a_43 | | | 16.94 | 27385 | 85 | - | - | abs |
| 225944.15-425918.9 | 290a-63 | PGC070317 | ESO290-36 | 16.55 | 14288 | 29 | 14683 | 10 | emi |
| 225956.84-430008.4 | 290b_159 | | | 16.05 | 26083 | 89 | - | - | emi |
| 230005.80-380741.8 | 346b_44 | | | 16.88 | 8519 | 22 | 8565 | 18 | emi |
| 230020.04-465354.0 | 290a_66 | | | 16.45 | 25896 | 75 | - | - | abs |
| 230026.77-400702.0 | 346a_45 | | | 16.92 | 8492 | 43 | 8454 | 24 | emi |
| 230029.80-413924.1 | 346b_79 | PGC070351 | ESO346-29 | 16.52 | 15169 | 79 | - | - | abs |
| 230050.58-443517.1 | 290a_68 | | | 16.16 | 20512 | 60 | 21000 | 5 | abs+emi |
| 230051.48-464808.9 | 290b_161 | PGC070364 | ESO290-40 | 12.79 | 5797 | 58 | 5885 | 4 | emi |
| 230053.12-440832.2 | 290a_69 | PGC070365 | ESO290-41 | 16.34 | 10299 | 15 | 10573 | 16 | emi |
| 230122.77-393156.0 | 346a_46 | | | 15.56 | 17212 | 94 | 17423 | 200 | emi |
| 230126.59-430026.7 | 290a_72 | PGC070391 | FAIR1032 | 15.62 | 15230 | 80 | - | - | abs |
| 230127.16-441610.3 | 290b_163 | PGC070390 | ESO290-43A | 15.57 | 16603 | 101 | 16741 | 57 | abs+emi |
| 230138.47-472516.3 | 290a_74 | | | 16.42 | 25576 | 109 | 25584 | 55 | emi |
| 230139.62-392332.9 | 346b_46 | | | 16.81 | 1311 | 67 | - | - | abs |
| 230212.45-375959.1 | 346b_68 | | | 16.93 | 10696 | 61 | 10831 | 4 | emi |
| 230218.23-460919.4 | 290a_77 | | | 16.50 | 5711 | 35 | 5795 | 8 | emi |
| 230229.02-404206.9 | 346b_47 | | | 15.80 | 17143 | 61 | - | - | abs |
| 230231.20-391859.2 | 346a_48 | | | 16.60 | 27342 | 134 | - | - | abs |
| 230233.40-441614.4 | 290a_78 | | | 15.60 | 15999 | 52 | 16238 | 23 | emi |
| 230252.20-420253.7 | 346b_71 | | | 16.41 | 20446 | 67 | - | - | abs |
| 230306.77-425808.3 | 290a_80 | | | 15.91 | 10616 | 32 | 10706 | 21 | emi |
| 230315.41-414640.9 | 346a_50 | | | 16.07 | 9974 | 64 | 10089 | 29 | emi |
| 230319.21-464927.0 | 290a_82 | PGC070476 | FAIR1034 | 15.57 | 10404 | 60 | - | - | abs+emi |
| 230330.12-403715.4 | 346b_50 | | | 16.93 | 8504 | 56 | - | - | abs |
| 230407.30-430959.2 | 290b_168 | PGC070494 | ESO290-51 | 14.20 | 12125 | 169 | 12468 | 10 | emi |
| 230423.74-375108.2 | 346a_51 | | | 16.93 | 13564 | 99 | - | - | emi |
| 230426.81-410948.2 | 346a_52 | PGC070514 | ESO346-31 | 12.34 | 9800 | 51 | 9941 | 14 | emi |
| 230445.12-400502.0 | 346b_52 | | | 16.46 | 17978 | 28 | 18012 | 24 | emi |
| 230507.12-444143.4 | 290b_268 | PGC070543 | ESO290-53 | 15.19 | 8668 | 82 | 8688 | 28 | emi |
| 230512.39-450020.9 | 290a_84 | | | 16.59 | 22198 | 71 | - | - | abs |
| 230531.31-380231.6 | 346b_67 | PGC070551 | ESO346-32 | 15.20 | 8669 | 84 | - | - | abs |
| 230539.32-465120.1 | 290a_85 | | | 16.55 | 27356 | 68 | 27166 | 123 | abs+emi |
| 230542.58-403443.0 | 346a_54 | | | 16.86 | 16869 | 88 | - | - | abs |
| 230553.66-374741.6 | 346a_55 | | | 16.71 | 17981 | 66 | - | - | abs |
| 230610.95-394820.2 | 346b_78 | | | 16.63 | 5434 | 32 | 5393 | 11 | emi |
| 230647.60-380051.5 | 346b_55 | | | 16.56 | 22720 | 161 | 22837 | 43 | emi |
| 230718.84-392434.6 | 346b_57 | | | 16.92 | 17630 | 112 | 17736 | 90 | emi |
| 230729.93-405519.8 | 346a_58 | | | 16.76 | 11525 | 66 | - | - | abs |
| 230730.37-382422.3 | 346b_58 | | | 16.87 | 17834 | 37 | 18142 | 27 | emi |
| 230757.59-392021.2 | 346a_59 | | | 16.53 | 18170 | 55 | - | - | abs |

Table 3: (continued)

| RA 1950 DEC<br>h m s d ' "<br>(1) (2) | field<br>number<br>(3) | PGC/LEDA<br>number<br>(4) | Alternat.<br>name<br>(5) | BT<br><br>(6) | Vcc<br>km/s<br>(7) | Err<br>km/s | Vem<br>km/s<br>(8) | Err<br>km/s | Notes<br><br>(9) |
|---|---|---|---|---|---|---|---|---|---|
| 230805.27-384418.7 | 346b_59 | | | 16.53 | 17014 | 36 | 16958 | 13 | emi |
| 230807.57-383459.9 | 346a_60 | PGC070630 | ESO346-35 | 15.52 | 2272 | 72 | 2238 | 113 | emi |
| 230843.33-405015.8 | 346a_61 | | | 16.65 | 25782 | 98 | 25918 | 30 | emi |
| 230852.11-393759.6 | 346b_75 | | | 16.56 | 17871 | 112 | - | - | abs |
| 230853.05-384726.3 | 346a_62 | | | 16.46 | 19336 | 76 | - | - | abs |
| 230853.57-421445.8 | 346b_62 | | | 16.63 | 16579 | 62 | 16506 | 31 | emi |
| 230903.99-403835.8 | 346a_63 | | | 16.23 | 11091 | 41 | 11081 | 8 | emi |
| 230905.44-405517.0 | 346b_63 | | | 16.72 | 16846 | 132 | - | - | abs |
| 230907.16-404129.4 | 346b_73 | | | 16.77 | 11183 | 30 | 11347 | 1 | emi |
| 230919.99-311056.9 | 470a_1 | | | 16.86 | 32638 | 69 | - | - | abs |
| 230928.01-284211.5 | 470a_2 | | | 16.67 | 9747 | 41 | - | - | abs |
| 230938.16-281441.7 | 470a_3 | | | 16.82 | 21266 | 45 | 21364 | 105 | emi |
| 230938.88-420603.4 | 346b_64 | | | 15.91 | 12313 | 49 | 12553 | 8 | emi |
| 230943.58-272623.4 | 470a_5 | | | 16.74 | 24977 | 68 | 25043 | 16 | emi |
| 230958.87-294640.0 | 470a_6 | | | 16.76 | 26430 | 59 | 26593 | 20 | emi |
| 231021.19-303528.1 | 470a_7 | | | 16.73 | 8921 | 54 | 9058 | 69 | emi |
| 231026.87-423019.8 | 346a_66 | | | 16.64 | 1224 | 12 | 1284 | 15 | emi |
| 231036.20-273415.6 | 470a_9 | | | 16.58 | 15935 | 88 | 15960 | 19 | emi |
| 231051.82-291725.7 | 470a_10 | | | 16.31 | 12232 | 91 | 12229 | 18 | emi |
| 231139.56-284924.0 | 470a_13 | | | 16.99 | 21169 | 54 | - | - | abs |
| 231200.48-303040.8 | 470a_14 | | | 16.48 | 6685 | 69 | 6614 | 22 | emi |
| 231221.36-282304.2 | 470a_15 | | | 16.31 | 25689 | 37 | 25760 | 15 | emi |
| 231229.63-275450.6 | 470a_16 | | | 16.61 | 8436 | 87 | 8499 | 7 | emi |
| 231232.69-315633.5 | 470a_17 | LEDA092795 | | 16.52 | 10362 | 172 | 10114 | 21 | emi |
| 231312.31-305158.9 | 470a_19 | | | 16.19 | 18670 | 162 | 19112 | 31 | emi |
| 231355.40-283127.7 | 470a_20 | | | 16.85 | 25082 | 107 | 25032 | 35 | emi |
| 231443.86-280802.1 | 470a_21 | | | 16.82 | 26343 | 47 | 26343 | 17 | emi |
| 231509.68-281043.5 | 470a_22 | | | 16.99 | 8850 | 21 | 8941 | 19 | emi |
| 231522.01-282654.3 | 470a_23 | | | 15.97 | 23846 | 106 | 23832 | 24 | emi |
| 231547.52-274521.1 | 470a_24 | | | 16.06 | 15824 | 51 | 15884 | 12 | emi |
| 231608.09-290756.7 | 470a_28 | | | 16.85 | 15605 | 93 | - | - | abs |
| 231614.20-290224.1 | 470a_28 | | | 16.92 | 6848 | 48 | 6861 | 17 | emi |
| 231653.77-294819.0 | 470a_29 | | | 16.94 | 15091 | 85 | - | - | abs |
| 231717.52-283408.9 | 470a_30 | | | 16.56 | 23333 | 100 | 23320 | 15 | emi |
| 231743.96-292233.2 | 470a_31 | | | 16.98 | 25377 | 48 | 25526 | 17 | emi |
| 231803.20-300813.6 | 470a_33 | | | 16.97 | 24897 | 61 | 24934 | 8 | emi |
| 231813.90-291350.4 | 470a_34 | | | 16.33 | 8614 | 89 | 8650 | 18 | emi |
| 231839.28-322809.4 | 470a_35 | | | 16.44 | 12158 | 36 | 12213 | 14 | emi |
| 231900.23-312442.4 | 470a_36 | | | 16.97 | 10824 | 53 | 10788 | 29 | emi |
| 231952.77-293318.0 | 470a_38 | PGC071245 | NGC7636 | 14.75 | 6797 | 65 | - | - | abs |
| 232037.56-290016.6 | 470a_39 | | | 16.99 | 32734 | 103 | 31080 | 15 | emi |
| 232059.66-292406.6 | 470a_40 | | | 16.95 | 15903 | 57 | - | - | abs |
| 232101.62-301521.5 | 470a_41 | | | 16.16 | 15331 | 75 | - | - | abs |
| 232116.15-301655.7 | 470a_42 | | | 16.89 | 15050 | 73 | 15096 | 12 | emi |
| 232148.66-293539.3 | 470a_43 | | | 15.51 | 6750 | 16 | 6831 | 15 | emi |
| 232218.73-322138.9 | 470a_44 | LEDA71377 | | 19.02 | 8573 | 34 | 8629 | 15 | emi |
| 232326.75-320733.8 | 470a_47 | LEDA71425 | | 15.64 | 18174 | 65 | - | - | abs |
| 232410.25-282841.2 | 470a_49 | | | 16.97 | 15707 | 75 | 15817 | 16 | emi |
| 232422.87-292208.7 | 470a_50 | | | 16.36 | 21249 | 81 | - | - | abs |
| 232427.27-292622.8 | 470a_51 | | | 16.96 | 20997 | 129 | - | - | abs+emi |
| 232458.23-304122.9 | 470a_52 | | | 16.99 | 10391 | 29 | 10456 | 2 | emi |
| 232514.17-315709.8 | 470a_53 | | | 16.31 | 18709 | 39 | 18817 | 3 | emi |
| 232516.27-293844.2 | 470a_54 | | | 16.74 | 15074 | 42 | 15027 | 16 | emi |
| 232545.17-292511.1 | 470a_55 | | | 16.92 | 20900 | 36 | - | - | abs |
| 232606.16-294916.2 | 470a_57 | PGC071551 | ESO470-9 | 15.38 | 15074 | 70 | - | - | abs |

Table 3: (continued)

| RA 1950 DEC h m s d ' " (1) (2) | field number (3) | PGC/LEDA number (4) | Alternat. name (5) | BT (6) | Vcc km/s (7) | Err km/s | Vem km/s (8) | Err km/s | Notes (9) |
|---|---|---|---|---|---|---|---|---|---|
| 232635.24-303600.0 | 470a_58 | | | 16.54 | 10385 | 53 | - | - | abs |
| 232647.31-313930.8 | 470a_59 | | | 16.35 | 10694 | 102 | 10671 | 38 | emi |
| 232652.89-312010.3 | 470a_60 | | | 16.38 | 6891 | 22 | 6640 | 42 | emi |
| 232656.29-290625.0 | 470a_61 | PGC071581 | IC5326 | 14.88 | 7259 | 34 | 7265 | 9 | emi |
| 232706.73-311208.2 | 470a_62 | | | 16.93 | 16194 | 62 | 15967 | 21 | emi |
| 232724.05-312558.3 | 470a_63 | PGC071601 | ESO470-13 | 15.32 | 10917 | 92 | 11139 | 22 | emi |
| 232738.05-300849.1 | 470a_64 | | | 16.22 | 15392 | 25 | 15462 | 34 | emi |
| 232754.27-290524.8 | 470a_65 | PGC071627 | ESO470-15 | 14.59 | 7204 | 46 | 7538 | 12 | emi |
| 232829.86-292656.5 | 470a_67 | | | 16.35 | 13772 | 20 | 13875 | 15 | emi |
| 232837.91-320832.6 | 470a-68 | | | 16.34 | 16015 | 39 | 16085 | 20 | emi |
| 232906.84-311051.1 | 470a_70 | | | 16.72 | 10863 | 38 | 10909 | 42 | emi |
| 232930.69-275529.2 | 470a_72 | | | 16.91 | 8251 | 20 | 8298 | 17 | emi |
| 232952.34-302006.9 | 470a_74 | | | 16.56 | 15499 | 59 | 15296 | 22 | emi |
| 233022.23-312626.3 | 470a_75 | | | 15.94 | 10693 | 48 | - | - | abs |
| 233045.61-285601.6 | 470a_76 | | | 16.83 | 19660 | 85 | 19748 | 8 | emi |
| 233057.73-290244.0 | 470a_77 | | | 16.71 | 14848 | 50 | 15042 | 38 | emi |
| 233100.77-295918.3 | 470a_78 | | | 16.47 | 15009 | 73 | - | - | abs |
| 233102.90-301327.2 | 470a_79 | | | 15.85 | 15210 | 75 | - | - | abs |
| 233140.82-322138.7 | 470a_80 | PGC071767 | MCG-5-55-25 | 15.25 | 13834 | 56 | - | - | abs |
| 233151.06-274713.8 | 470a_81 | | | 16.71 | 26163 | 88 | - | - | abs |
| 233210.17-300546.6 | 470a_82 | | | 16.03 | 15167 | 28 | 15145 | 11 | emi |
| 233240.62-291503.6 | 470a_83 | | | 16.65 | 15199 | 22 | 15184 | 16 | emi |
| 233304.43-303641.1 | 470a_84 | | | 16.94 | 9081 | 81 | 9244 | 40 | emi |
| 233305.89-302119.1 | 470a_85 | | | 16.78 | 10585 | 70 | 10588 | 17 | emi |
| 233308.45-313521.1 | 470a_86 | | | 16.71 | 13122 | 71 | 13230 | 10 | emi |
| 233328.81-324656.9 | 470a_87 | | | 16.69 | 15560 | 83 | - | - | abs |
| 233338.16-315247.4 | 470a_88 | PGC071871 | | 14.87 | 18752 | 49 | - | - | abs |
| 233625.56-255649.6 | 537a_3 | PGC072012 | ESO536-14 | 14.78 | 9459 | 45 | 9452 | 3 | emi |
| 233655.88-253234.8 | 537a_4 | | | 16.77 | 14650 | 49 | 14719 | 7 | emi |
| 233713.88-230200.8 | 537a_6 | | | 15.74 | 7808 | 29 | 7911 | 15 | emi |
| 233723.95-271933.9 | 537a_7 | | | 16.39 | 19539 | 79 | 19306 | 13 | emi |
| 233735.12-230031.8 | 537a_8 | | | 16.57 | 7946 | 46 | 7997 | 15 | emi |
| 233755.16-264958.6 | 537a_9 | | | 16.29 | 15052 | 59 | 14860 | 20 | emi |
| 233834.71-225826.2 | 537a_10 | | | 16.94 | 14366 | 56 | - | - | abs |
| 233842.73-251847.4 | 537a_11 | | | 16.71 | 15611 | 25 | 15640 | 17 | emi |
| 233903.62-253437.0 | 537a_13 | | | 16.06 | 16456 | 129 | 16764 | 8 | emi |
| 233956.30-271150.4 | 537a_14 | | | 16.63 | 19235 | 73 | 20098 | 2 | emi |
| 234041.77-252223.9 | 537a_15 | | | 16.73 | 15936 | 29 | 16012 | 3 | emi |
| 234053.73-262104.7 | 537a_16 | LEDA89427 | | 16.25 | 15969 | 25 | 15895 | 1 | emi |
| 234106.20-231921.2 | 537a_17 | | | 16.68 | 9192 | 44 | - | - | abs |
| 234136.42-244142.4 | 537a_18 | | | 16.64 | 6703 | 20 | 6730 | 15 | emi |
| 234139.33-262500.7 | 537a_19 | | | 16.94 | 16380 | 39 | - | - | abs |
| 234147.24-241554,8 | 537a_20 | PGC072264 | ESO537-2 | 15.27 | 14355 | 49 | - | - | abs |
| 234155.93-262828.3 | 537a_21 | | | 16.67 | 16427 | 27 | 16474 | 15 | emi |
| 234201.11-253029.5 | 537a_22 | | | 16.48 | 14711 | 41 | 14778 | 13 | emi |
| 234202.41-240517.1 | 537a_23 | PGC072282 | ESO537-3 | 15.33 | 14772 | 83 | - | - | abs |
| 234234.43-240517.1 | 537a_26 | PGC072310 | ESO537-4 | 16.74 | 9817 | 55 | 9814 | 16 | emi |
| 234249.50-271036.8 | 537a_28 | | | 15.78 | 14488 | 115 | - | - | abs |
| 234252.94-245311.9 | 537a_29 | | | 16.36 | 9835 | 41 | - | - | abs |
| 234300.11-271758.8 | 537a_30 | | | 16.61 | 15068 | 67 | - | - | abs |
| 234305.94-252713.7 | 537a_31 | | | 16.61 | 16447 | 29 | 16975 | 13 | emi |
| 234316.60-240033.2 | 537a_32 | PGC072342 | ESO537-7 | 11.46 | 9924 | 82 | 10257 | 2 | emi |
| 234328.95-231035.2 | 537a_33 | PGC072349 | ESO537-8 | 16.45 | 14275 | 55 | - | - | abs+emi |
| 234334.98-252432.0 | 537a_34 | | | 16.55 | 9930 | 65 | - | - | abs |

Table 3: (continued)

| RA 1950 DEC<br>h m s d ' "<br>(1) (2) | field<br>number<br>(3) | PGC/LEDA<br>number<br>(4) | Alternat.<br>name<br>(5) | BT<br><br>(6) | Vcc<br>km/s<br>(7) | Err<br>km/s | Vem<br>km/s<br>(8) | Err<br>km/s | Notes<br><br>(9) |
|---|---|---|---|---|---|---|---|---|---|
| 234340.60-260408.5 | 537a_35 | PGC072356 | ESO537-9 | 16.27 | 8648 | 47 | 8632 | 15 | emi |
| 234343.74-250447.6 | 537a_36 | LEDA89431 | | 16.66 | 14769 | 43 | 14455 | 16 | emi |
| 234352.09-253829.6 | 537a_37 | | | 16.61 | 10063 | 75 | - | - | abs |
| 234356.74-273917.1 | 537a_38 | | | 16.72 | 15022 | 24 | 15116 | 15 | emi |
| 234357.69-231032.1 | 537a_39 | | | 16.56 | 8525 | 16 | 8588 | 15 | emi |
| 234413.84-260038.9 | 537a_40 | | | 16.38 | 9888 | 56 | - | - | abs |
| 234424.35-231135.0 | 537a_42 | LEDA089433 | | 16.19 | 17456 | 22 | 17546 | 15 | emi |
| 234437.41-240500.8 | 537a_43 | | | 16.41 | 22127 | 22 | 22091 | 15 | emi |
| 234451.67-261346.0 | 537a_44 | | | 16.43 | 9885 | 70 | - | - | abs |
| 234454.36-275623.8 | 537a_45 | LEDA085761 | DRCG54-91 | 15.88 | 8605 | 58 | - | - | abs |
| 234503.05-260533.2 | 537a_46 | | | 16.81 | 16619 | 40 | 16625 | 18 | emi |
| 234507.95-252632.1 | 537a_47 | | | 16.79 | 16385 | 78 | - | - | abs |
| 234518.75-275822.2 | 537a_48 | LEDA085772 | DRCG54-90 | 13.35 | 15035 | 22 | 15007 | 17 | emi |
| 234555.12-261913.4 | 537a_49 | | | 16.45 | 23766 | 53 | - | - | abs |
| 234612.35-275237.2 | 537a_50 | LEDA085791 | DRCG54-94 | 16.69 | 19581 | 74 | - | - | abs |
| 234619.91-221809.9 | 537a_51 | PGC072497 | NGC7758 | 15.98 | 13084 | 50 | - | - | abs |
| 234624.75-233118.3 | 537a_52 | | | 16.91 | 17142 | 50 | 17105 | 15 | emi |
| 234647.70-241948.2 | 537a_53 | PGC072524 | ESO537-13 | 12.59 | 14871 | 19 | 14839 | 15 | emi |
| 234710.38-275845.0 | 537a_54 | LEDA085810 | DRCG54-89 | 16.57 | 19163 | 51 | 19332 | 15 | emi |
| 234720.16-241813.4 | 537a_55 | | | 16.16 | 17218 | 60 | - | - | abs |
| 234722.97-270809.6 | 537a_56 | PGC072550 | ESO537-15 | 15.99 | 10070 | 50 | - | - | abs |
| 234752.39-243222.7 | 537a_57 | | | 16.34 | 14903 | 54 | - | - | abs |
| 234848.07-243216.4 | 537a_58 | | | 16.65 | 15567 | 37 | 15493 | 7 | emi |
| 234907.63-260739.9 | 537a_59 | PGC072650 | ESO472-1 | 16.14 | 3466 | 57 | 3701 | 17 | emi |
| 234929.12-260812.5 | 537a_60 | PGC072676 | ESO472-2 | 16.01 | 3054 | 32 | 3236 | 12 | emi |
| 234932.29-233844.1 | 537a_61 | | | 16.67 | 19808 | 60 | - | - | abs |
| 235003.45-240633.7 | 537a_62 | | | 16.12 | 15381 | 113 | - | - | abs |
| 235045.23-253749.4 | 537a_63 | | | 15.75 | 9976 | 47 | 10083 | 15 | emi |
| 235057.61-233019.8 | 537a_64 | | | 16.43 | 19883 | 46 | - | - | abs |
| 235107.04-234009.7 | 537a_65 | | | 16.61 | 15008 | 71 | 15052 | 20 | emi |
| 235135.22-225343.6 | 537a_66 | | | 16.75 | 15054 | 58 | 15270 | 18 | emi |
| 235152.84-274757.8 | 537a_67 | PGC072827 | ESO471-40 | 16.33 | 16526 | 97 | - | - | abs |
| 235205.52-254600.7 | 537a_68 | | | 16.56 | 3006 | 15 | 3043 | 15 | emi |
| 235210.35-255704.6 | 537a_69 | PGC072841 | ESO472-7 | 15.64 | 9071 | 101 | 9091 | 17 | emi |
| 235239.90-240636.9 | 537a_70 | PGC072860 | ESO472-10 | 14.42 | 3879 | 29 | 3962 | 16 | emi |
| 235335.27-274443.8 | 537a_72 | PGC072933 | ESO471-44 | 15.51 | 3007 | 10 | 3033 | 15 | emi |
| 235357.12-241302.2 | 537a_73 | | | 16.62 | 22411 | 72 | 22399 | 17 | emi |
| 235439.34-241818.1 | 537a_75 | | | 16.79 | 15922 | 124 | 15471 | 17 | emi |
| 235458.48-251216.9 | 537a_76 | | | 16.28 | 19289 | 30 | 19351 | 15 | emi |
| 235522.57-225944.8 | 537a_77 | | | 16.64 | 15612 | 41 | 15387 | 6 | emi |
| 235645.74-244315.5 | 537a_79 | PGC073155 | ESO472-12 | 12.99 | 15493 | 47 | 15647 | 16 | emi |
| 235721.40-273045.0 | 537a_80 | PGC073192 | MCG-5-1-21 | 15.52 | 8188 | 33 | - | - | abs |
| 235740.23-252754.4 | 537a_81 | | | 16.67 | 25403 | 109 | - | - | abs |
| 235740.41-270032.7 | 537a_82 | | | 16.01 | 17945 | 37 | - | - | abs |
| 235751.18-261922.0 | 537a_83 | | | 16.63 | 15174 | 53 | - | - | abs |
| 235821.95-261102.4 | 537a_85 | | | 16.39 | 15065 | 41 | - | - | abs |
| 235837.06-271755.1 | 537a_86 | | | 16.35 | 8072 | 80 | 8106 | 17 | emi |
| 235848.16-252826.0 | 537a_87 | | | 16.53 | 8264 | 31 | 8305 | 4 | emi |
| 235922.68-243335.3 | 537a_88 | | | 16.62 | 19737 | 59 | 19910 | 16 | emi |
| 235956.65-273428.2 | 537a_89 | | | 16.39 | 8405 | 47 | - | - | abs |
| 224907.64-451306.7 | 290a_37 | | | 15.66 | 16272 | 81 | - | - | emi |
| 225253.11-423049.6 | 290a_46 | | | 15.69 | 16455 | 112 | - | - | abs |
| 225536.18-425446.0 | 290a_54 | | | 16.39 | 24563 | 209 | - | - | abs+emi |
| 230057.93-433851.2 | 290a_70 | | | 16.26 | 10298 | 110 | - | - | emi |

Table 3: (continued)

| RA 1950 DEC h m s d ' " (1) (2) | field number (3) | PGC/LEDA number (4) | Alternat. name (5) | BT (6) | Vcc km/s (7) | Err km/s | Vem km/s (8) | Err km/s | Notes (9) |
|---|---|---|---|---|---|---|---|---|---|
| 223436.88-462438.1 | 290b_102 | | | 16.64 | 22297 | 61 | - | - | abs |
| 224854.38-465122.4 | 290b_140 | | | 16.86 | 17250 | 94 | - | - | : abs |
| 225235.62-434501.1 | 290b_145 | | | 16.90 | 22331 | 69 | - | - | : abs |
| 231552.16-301437.7 | 470a_25 | | | 16.89 | 35530 | 51 | - | - | abs |
| 232343.62-300501.9 | 470a_48 | | | 15.79 | 19263 | 66 | - | - | abs |
| 232900.98-325302.1 | 470a_71 | PGC0071670 | ESO408-7 | 15.77 | 16375 | 32 | - | - | emi |
| 233357.13-314013.0 | 470a_90 | | | 16.91 | 19896 | 87 | - | - | abs |
| 235625.27-233615.2 | 537a_78 | | | 16.57 | 32558 | 72 | - | - | abs |
| 235755.64-274005.7 | 537a_84 | | | 16.89 | 13915 | 56 | - | - | abs |
| 235928.98-273157.2 | 537a_89 | | | 16.70 | 14299 | 95 | - | - | abs |
| 224641.29-374500.7 | 346b_13 | LEDA69800 | | 16.84 | 8737 | 52 | - | - | abs |
| 225639.21-385927.7 | 346b_34 | | | 16.93 | 16150 | 70 | - | - | emi |
| 224315.27-420427.4 | 346b_80 | | | 16.98 | 6079 | 63 | - | - | emi |
| 225829.88-384525.9 | 346b_90 | | | 16.70 | 15989 | 82 | - | - | emi |
| 225725.28-401832.0 | 346a_37 | | | 16.68 | 27170 | 166 | - | - | abs |
| 231030.08-392516.3 | 346a_700 | | | 16.53 | 19278 | 71 | - | - | : abs |
| 224341.33-393252.8 | 346b_2 | | | 16.98 | 23702 | 127 | - | - | emi |
| 224541.34-401302.5 | 346b_8 | | | 16.45 | 2381 | 74 | - | - | emi |
| 225313.90-380320.5 | 346b_28 | | | 16.93 | 25193 | 60 | - | - | abs |
| 225600.77-423943.9 | 346b_33 | | | 16.42 | 19487 | 55 | - | - | : abs |
| 225747.04-421705.2 | 346b_37 | | | 16.50 | 15762 | 107 | - | - | abs |
| 230118.23-411552.6 | 346b_45 | | | 16.95 | 9352 | 72 | - | - | : abs |
| 230836.66-424651.9 | 346b_60 | | | 17.00 | 5502 | 87 | - | - | : abs |
| 225051.35-395654.5 | 346b_65 | | | 16.67 | 10471 | 57 | - | - | : abs |
| 231027.49-423323.5 | 346b_85 | | | 16.71 | 15143 | 63 | - | - | : abs |
| 230322.70-373916.7 | 346b_91 | | | 16.84 | 11808 | 85 | - | - | abs |
| 224246.17-401722.0 | 346b_96 | LEDA69667 | | 16.81 | 5531 | 93 | - | - | : abs |
| 224325.72-380913.8 | 346a-2 | LEDA85304 | | 16.27 | 30421 | 83 | - | - | : abs |
| 224351.04-413949.4 | 346a_3 | | | 16.87 | 17258 | 92 | - | - | abs |
| 224424.69-373755.7 | 346a_4 | | | 16.71 | 28788 | 109 | - | - | abs+emi |
| 224431.27-410847.2 | 346a_5 | LEDA85307 | | 16.57 | 19960 | 149 | - | - | abs |
| 224453.86-373842.1 | 346a_6 | | | 16.71 | 39006 | 75 | - | - | : abs |
| 224927.31-410128.0 | 346a_20 | | | 16.75 | 19909 | 159 | - | - | : abs |
| 225655.61-413516.0 | 346a_35 | | | 16.80 | 8168 | 56 | - | - | : abs |
| 225905.75-384549.8 | 346a_40 | LEDA95165 | | 16.46 | 3799 | 104 | - | - | : emi |
| 225947.50-422755.3 | 346a_44 | | | 16.41 | 17776 | 76 | - | - | abs+emi |
| 230203.27-412533.5 | 346a_47 | | | 16.70 | 19928 | 125 | - | - | abs |
| 231305.70-403359.5 | 346a_300 | LEDA70869 | | 15.30 | 15688 | 70 | - | - | : abs |
| 225157.74-433023.0 | 290a_55 | | | 16.13 | 39678 | 68 | - | - | abs |
| 225932.57-460721.7 | 290a_62 | | | 16.87 | 22748 | 72 | - | - | : abs |
| 223848.59-424234.0 | 290b_110 | | | 16.38 | 12093 | 149 | - | - | : emi |
| 224510.05-423259.5 | 290b_126 | | | 16.69 | 20543 | 64 | - | - | : abs |
| 224654.69-471452.6 | 290b_134 | | | 16.50 | 33059 | 105 | - | - | abs |
| 225719.05-465628.1 | 290b_154 | | | 16.92 | 25180 | 78 | - | - | : abs |
| 224119.08-452304.8 | 290b_215 | LEDA69621 | | 17.24 | 32552 | 83 | - | - | : abs |
| 224516.82-461959.9 | 290b_227 | | | 16.29 | 15401 | 138 | - | - | : abs |
| 225341.41-472940.1 | 290b_246 | | | 16.78 | 4339 | 83 | - | - | : abs |
| 225606.30-461544.0 | 290b_251 | | | 16.82 | 22051 | 55 | - | - | : abs |
| 000123.74-232914.1 | 537a_100 | | | 15.50 | 14388 | 66 | - | - | : abs |

Note: the ':' in the notes column indicates a poor determination of the redshift.
'abs' denotes redshift measured from absorption lines
'emi' denotes redshift measured from emission lines